# MIGRATION OF DUST PARTICLES TO THE TERRESTRIAL PLANETS

S.I. Ipatov [(1)], J.C. Mather [(2)]

[(1)]University of Maryland, College Park, MD 20742, USA; Space Research Institute, Moscow, Russia, Email: sipatov@umd.edu. http://www.astro.umd.edu/~ipatov
[(2)] NASA/Goddard Space Flight Center, Greenbelt, MD, 20771, USA, Email: john.c.mather@nasa.gov

**ABSTRACT/RESUME**

The orbital evolution of asteroidal, trans-Neptunian, and cometary dust particles under the gravitational influence of planets, the Poynting-Robertson drag, radiation pressure, and solar wind drag was integrated. Results of our runs were compared with the spacecraft observations of the number density of dust particles and with the WHAM observations of velocities of zodiacal particles. This comparison shows that the fraction of cometary dust particles of the overall dust population inside Saturn's orbit is significant and can be dominant. The probability of a collision of an asteroidal or cometary dust particle with the Earth during its dynamical lifetime is maximum at diameter $d\sim 100$ μm.

## 1. INTRODUCTION

There are a lot of papers on migration of dust (e.g., [1-12], see more references and comparison of our runs with previous results in [13-14]). In this paper we summarize our previous studies [13-14] based on a larger number of integrations than earlier. Particular attention is paying to the probabilities of collisions of dust particles with the terrestrial planets. In contrast to papers by other scientists, we studied the orbital evolution of dust particles and probabilities of their collisions with the Earth for a wider range of diameters (up to 1000 μm and greater) of asteroidal and cometary particles and considered also migration of dust particles produced by comets 10P and 39P and by long-period comets.

## 2. MODEL

We integrated [13-14] the orbital evolution of about 15,000 asteroidal, cometary, and trans-Neptunian dust particles under the gravitational influence of planets, the Poynting-Robertson drag, radiation pressure, and solar wind drag, varying the values of the ratio β between the radiation pressure force and the gravitational force from ≤0.0004 to 0.4 (for silicates, such values of β correspond to particle diameters $d$ between ≥1000 and 1 microns; $d$ is proportional to 1/β). The considered cometary particles started from comets 2P/Encke ($a$≈2.2 AU, $e$≈0.85, $i$≈12°), 10P/Tempel 2 ($a$≈3.1 AU, $e$≈0.526, $i$≈12°), 39P/Oterma ($a$≈7.25 AU, $e$≈0.246, $i$≈2°), and from long-period comets ($e$=0.995, $q=a(1-e)$=0.9 AU or $q$=0.1 AU, $i$ was distributed between 0 and 180°, particles started at perihelion). 10P and 39P are examples of typical Jupiter-family comets located inside and outside Jupiter's orbit, respectively. The integration continued until all of the particles either collided with the Sun or reached 2000 AU from the Sun. For small β, considered times exceeded 50-80 Myr (240 Myr for trans-Neptunian particles). In our runs orbital elements were stored with a step of $d_t$ of ≤20 yr for asteroidal and cometary particles and 100 yr for trans-Neptunian particles. The planets were assumed to be material points; however, using orbital elements obtained with a step $d_t$, we calculated the mean collision probability of a particle with a planet during a dynamical lifetime of the particle (destruction of particles at their mutual collisions was not considered) $P=P_\Sigma/N$, where $P_\Sigma$ is the probability for all $N$ considered particles in one run.

## 3. PROBABILITIES OF COLLISIONS OF DUST WITH THE TERRESTRIAL PLANETS

The probabilities $P$ of collisions of dust particles with the Earth versus β are presented in Fig. 1.

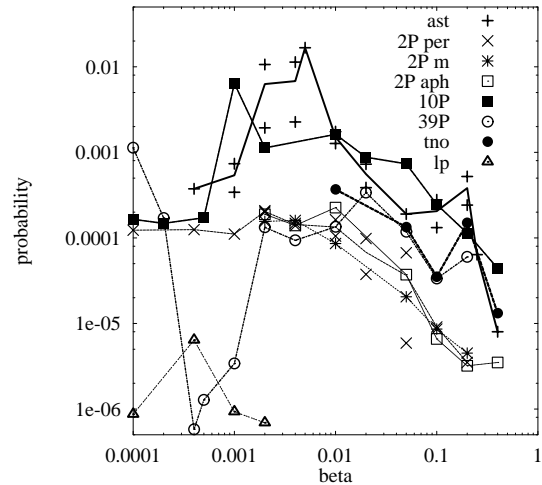

Figure 1. The probabilities $P$ of collisions of dust particles with the Earth versus β for particles started from asteroids (ast), trans-Neptunian objects (tno), Comet 2P at perihelion (2P per), Comet 2P at aphelion (2P aph), Comet 2P in the middle between perihelion and aphelion (2P m), Comet 10P (10P), Comet 39P (39P), and long-period comets (LP) at $e$=0.995 and $q$=0.9 AU. If there are two points for the same β, then a plot is drawn via their mean value.

For some values of β, we considered two runs with slightly different initial data [14]. The obtained values of $P$ can differ by a factor of several for such two runs because the number $N$ of particles in each run was small (100-250), and a few objects can get orbits (e.g., with small inclinations) with relatively high probability of a collision with the Earth.

The probability $P$ of a collision of an asteroidal dust particle with the Earth was found to have a maximum (~0.001-0.02) at 0.002≤β≤0.01, i.e., at $d$~100 μm. This is in accordance with cratering records in the lunar soil and also with particles record on the panels of the Long Duration Exposure Facility, which showed that the mass distribution of dust particles encountering Earth peaks at $d$=200 μm [5, 15]. Grün et al. [15] noted that at 1 AU the collisional destruction rate of particles with masses $m≥10^{-3}$g is about 10 times larger than the rate of collisional production of fragment particles in the same size. The probabilities $P$ of collisions of asteroidal particles with Venus did not differ much from those for Earth, whereas for Mars they were by an order of magnitude smaller at β≥0.01 compared to Earth, and were nearly similar to those for Earth at β~0.0004-0.001 [14]. We suppose that smaller values of $P$ at smaller β<0.005 in Fig. 1 are caused by the fact that particles with $d$~100 μm more often have less inclined and less eccentric Earth-crossing orbits, which cause higher probability of collisions, than more massive particles.

The probability $P$ of a collision of a particle started from Comet 10P with a terrestrial planet sometimes differed by a factor of several from that for an asteroidal particle of the same size. In turn, for Comet 2P dust debris, the $P$ values were found usually smaller than for asteroidal and 10P particles: for Earth at 0.002≤β≤0.01, $P$ was by an order of magnitude (and sometimes even more) smaller for 2P particles than for asteroidal particles. So the fraction of particles started from high-eccentricity comets such as Comet 2P (among particles from different sources) is much smaller for particles collided with the Earth than for particles moving in near-Earth space. For 2P particles at some β, $P$ is by a factor of 2 or 4 greater for Venus than for Earth [14].

For trans-Neptunian and 39P particles, maximum values of the probability of their collisions with the Sun (0.2-0.3) were reached at 0.05≤β≤0.1. For β≥0.05, the fraction of trans-Neptunian particles collided with the Sun was less than that of asteroidal particles by a factor of 4-6. At 0.01≤β≤0.2, the probabilities of collisions of trans-Neptunian particles with Earth and Venus were ~(0.3-4)·$10^{-4}$ and were usually less than those for asteroidal particles by a factor of less than 4. The ratio of values of time $T$ during which a particle has a perihelion distance $q$<1 AU for asteroidal particles to the values of $T$ for trans-Neptunian particles was about 3-7 at β≥0.1 and about 20 at β=0.05. The mean values $e_m$ and $i_m$ of eccentricities and inclinations at distance $R$=1 AU from the Sun were mainly greater for trans-Neptunian particles than those for asteroidal particles. Nevertheless, the ratio $P/T$ was greater for trans-Neptunian particles. It may be caused by that perihelia or aphelia of migrating trans-Neptunian particles more often were close to the orbit of the Earth, or the fraction of Earth-crossing trans-Neptunian particles with small $e$ and $i$ was greater (though $e_m$ and $i_m$ were greater) than for asteroidal particles.

At β=0.0001 one 39P particle moved in an Earth-crossing orbit located inside Jupiter's orbit for 6 Myr, and due to this particle, the values of $P$ and $T$ for this run were much greater than those for other 39P runs. For 39P particles at β≤0.001, one need to consider many thousands of particles in order to get reliable statistics because for such runs the probability of a collision of one particle with a terrestrial planet can be greater than the total probability of collisions of thousands other particles. Comet 39P is located outside Jupiter's orbit, and studies of the orbital evolution of dust particles produced by this comet help to better understand migration of trans-Neptunian particles to the terrestrial planets at small β. At 0.01≤β≤0.2 the values of $P$ for trans-Neptunian particles were similar to those for 39P particles (~$10^{-4}$), but the times in Earth-crossing orbits for trans-Neptunian particles were smaller by a factor of several than those for 39P particles. Due to a small fraction of large ($d$>1000 μm) particles that can move in Earth-crossing orbits for a long time, it may be possible that the probability of a collision of such trans-Neptunian particles with the Earth can be of the same order of magnitude as that for $d$<50 μm, but much more runs are needed for accurate estimates.

We also considered migration of particles started at perihelion from long-period comets with $q$<1 AU, $e$=0.995, and $i$ distributed between 0 and 180°. At β≥0.004, all particles reached 2000 AU at $t$≤5 Kyr. At β≤0.002 for $q$=0.9 AU and at β≤0.0004 for $q$=0.1 AU, dynamical lifetimes of some particles exceeded several Myr. So only relatively large ($d$>100 μm) particles started from near-parabolic comets collided with the Earth ($P$≈2·$10^{-5}$ for $q$=0.1 AU at β=0.0004).

Interstellar particles can be effective in destruction of trans-Neptunian dust particles through collisions, especially for 9≤$d$≤50 μm, as it is argued in [7]. Larger particles may survive because interstellar grains are too small to destroy them in a single impact. Since the total mass of the trans-Neptunian belt exceeds that of the asteroid belt by two orders of magnitude (or even more), and the derived in our model mean residence times ratio in orbits with $q$<1 AU for asteroid and trans-Neptunian particles is less than 20 at β≥0.05, then for $d$~1-10 μm the fraction of non-icy trans-Neptunian dust of the overall dust population can be significant even at $R<3$ AU.

## 4. DISTRIBUTION OF MIGRATING DUST OVER DISTANCE FROM THE SUN

Based on the above runs, we studied [14] the distribution of number density $n_s$ (i.e., the number of particles per unit of volume) near ecliptic over distance $R$ from the Sun. The plot of $n_s$ versus $R$ at β=0.05 is presented in Fig. 2. For asteroidal particles, $n_s$ quickly decreases with an increase of $R$. For β=0.2, $n_s$ was smaller at $R$=5 AU than at $R$=3 AU by a factor of 8, 5, and 8 for asteroidal, 2P, and 10P particles, respectively. For β<0.05 such particles almost never reach 5 AU, and the above factor is much greater. So dust particles originated from small bodies located inside Jupiter's orbit can not explain the constant number density of dust particles at $R$~3-18 AU, which was observed during the flights of Pioneer 10 and 11. At $R$>5 AU, many of the dust particles could have come from bodies moving beyond Jupiter's orbit.

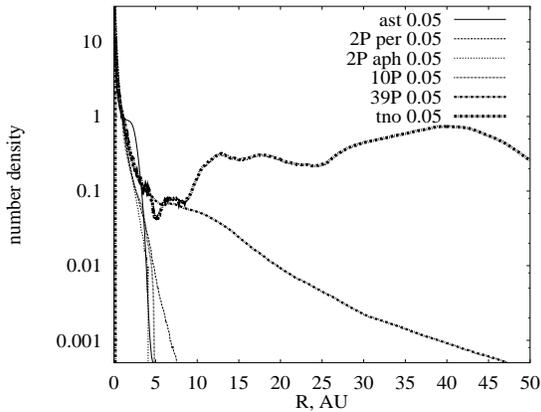

Figure 2. The number density of migrating dust particles over their distance $R$ from the Sun at β=0.05 for particles started from asteroids (ast), trans-Neptunian objects (tno), Comet 2P at perihelion (2P per), Comet 2P at aphelion (2P aph), Comet 10 P, and Comet 39P. Number density at 1 AU is considered to be equal to 1.

In our runs at β≥0.05, the number density $n_s$ of considered trans-Neptunian particles near ecliptic at $R$=1 AU was greater than at $R$>1 AU. At 0.1≤β≤0.4 and 2<$R$<45 AU (at β=0.05 for 11<$R$<50 AU) for trans-Neptunian particles, $n_s$ varied with $R$ by less than a factor of 4, but at $R$=5 AU it was smaller by at least a factor of 2 than at 15 AU. The number density of trans-Neptunian dust particles is smaller by a factor of several at 5<$R$<10 AU than at $R$=18 AU, so the fraction of cometary dust particles at $R$~5-10 AU is considerable. Similar conclusions were made in the previous studies of number density vs. $R$ (e.g., [12, 16]) for other values of β and other sources of cometary particles.

## 5. VELOCITIES OF DUST PARTICLES

We studied [14, 17-18] how the solar spectrum is changed by scattering by dust particles. Positions of particles were taken from the runs of migration of dust particles. For each such stored position, we calculated many (~$10^2$-$10^4$ depending on a run) different positions of a particle and the Earth during the period $P_{rev}$ of revolution of the particle around the Sun, considering that orbital elements do not vary during $P_{rev}$. Three different scattering functions were considered [14, 17]. For each considered position, we calculated velocities of a dust particle relative to the Sun and the Earth and used these velocities and the scattering function for construction of the solar spectrum received at the Earth after been scattering by different particles located at some beam (line of sight) from the Earth. The direction of the beam is characterized by elongation ε and inclination $i$. Particles in the cone of 2.5° around this direction were considered. All positions of particles during their dynamical lifetimes obtained in a single run (with fixed β and the same source of particles) were used for construction of one plot.

Plots of velocities of Mg I line (at $i$=0) versus elongation ε (measured eastward from the Sun), were compared [14, 18-19] with the WHAM observational plots presented in [20]. Such comparison with the WHAM observations was not made by other scientists. The plots obtained for different considered scattering functions were similar at 30°<ε<330°, the difference was greater for more close direction to the Sun. For future observations of velocities of the zodiacal light, it is important to pay particular attention to ε between 90° and 120°. For these values of ε, the difference between different plots for different sources of dust was maximum. In our opinion, the main conclusion of the comparison of the 'velocity-elongation' plots obtained at observations with those for our models is that asteroidal dust doesn't dominate in the zodiacal light and a lot of zodiacal dust particles were produced by comets, including high eccentricity comets. This result is in agreement with our studies of dynamics of Jupiter-family comets [21-23], which showed that there could be a lot of extinct comets moving in orbits with high eccentricities inside Jupiter's orbit. Significant contribution of cometary dust to the zodiacal cloud was considered by several other authors (e.g., in [24] it was supposed to be ~75%).

The values of the 'full width at half maximum' (FWHM), i.e., the width at (min+max)/2, for a plot of the intensity of light vs. its wavelength near Mg I line presented in Fig. 5 in [18] are mainly greater than the FWHM obtained in our runs for asteroidal dust, but are mainly smaller than the FWHM for particles started from Comet 2P and long-period comets. Such width characterizes the scatter in velocities. For particles started from asteroids, comets 2P, 10P and 39P, and long-period comets, the mean values of FWHM are about 74, 81-88, 76-77, 76-77, 73-86 km/s, respectively (the observational value is equal to 77).

## 6. CONCLUSIONS

The probabilities of collisions of migrating asteroidal and cometary dust particles with the terrestrial planets during the dynamical lifetimes of these particles were maximum at diameter $d \sim 100$ μm, which is in accordance with the analysis of microcraters.

Cometary dust particles (produced both inside and outside Jupiter's orbit) are needed to explain the constant number density of particles at 3-18 AU. Comparison of the velocities of particles obtained in our runs with the velocities of zodiacal particles obtained at the WHAM observations shows that only asteroidal dust particles cannot explain these observations, and particles produced by comets, including high-eccentricity comets, are needed for such explanation.